\documentclass[12pt]{article}

\usepackage{amsmath,amsthm,bm}

\usepackage{amsfonts}
\usepackage{natbib}

\usepackage[dvips]{graphicx}

\makeatletter
\newcommand{\@makemycaption}[2]{%
\vspace{10pt}%
{\textbf{#1}:#2\par}%
}
\renewcommand{\figure}{%
\let\@makecaption\@makemycaption\@float{figure}}
\renewcommand{\table}{%
\let\@makecaption\@makemycaption\@float{table}}
\makeatother

\date{}

\newcommand{\wt}[1]{\widetilde{#1}}
\newcommand{\what}[1]{\widehat{#1}}

\numberwithin{equation}{section}

\begin{document}

\begin{center}
  \textbf{\large Multistage tests of multiple hypotheses}

\bigskip

\textbf{Jay Bartroff}\\ Department of Mathematics, University of Southern California, Los
 Angeles, CA, USA {\renewcommand{\thefootnote}{}\footnote{Address
 correspondence to  Jay Bartroff, Department of Mathematics, University of Southern
California, 3620 South Vermont Ave., KAP 108, Los Angeles, CA 90089-2532, USA; E-mail: bartroff@usc.edu}}
\medskip

\textbf{Tze Leung Lai}

Department of Statistics, Stanford University, Stanford, CA, USA
\end{center}

\bigskip

\noindent\textbf{\small Abstract:} {\small Conventional multiple hypothesis tests use step-up, step-down, or closed testing methods to control the overall error rates.  We will discuss marrying these methods with adaptive multistage sampling rules and stopping rules to perform efficient multiple hypothesis testing in sequential experimental designs.  The result is a multistage step-down procedure that adaptively tests multiple hypotheses while preserving the family-wise error rate, and extends Holm's (1979) step-down procedure to the sequential setting, yielding substantial savings in sample size with small loss in power. 
\vspace{.5in}



\newpage

\section{\large Introduction}

Multiple testing problems of the sequential or, more generally, multistage nature occur frequently in statistics. For example, in sequential fault detection and diagnosis \citep{Nikiforov95,Lai00}, after detecting that a change in the system has occurred at some time point, the task is to isolate this changepoint to one of $k$ time intervals or diagnose  it as one of $k$ change types. Another area rich with examples is sequential clinical trials with multiple endpoints \citep[e.g.,][]{Tang93,Tang99} in which patients are accrued, treated, and evaluated sequentially with regard to~$k$ different features, none of which may have priority. Some recent new areas of applications are quantitative finance, in empirical tests of the profitability of trading strategies~\citep{Romano05b}, and genomics \citep{Ge03}.

Based on data coming from a parametric family $F_\theta$, $\theta\in\Theta$, of distributions, we will be concerned with testing a set of hypotheses $H_1,\ldots,H_k\subseteq\Theta$. A hypothesis $H_i$ is \emph{true} if the true~$\theta$ lies in $H_i$. If $I\subseteq\{1,\ldots,k\}$ is the set of indices of the true hypotheses, then the \emph{family-wise error rate~(FWE)} is defined as the probability
\begin{equation}
\label{eq:FWE}P(\mbox{some $H_i$, $i\in I$, is rejected}).
\end{equation} Other and somewhat less stringent notions of error rate have been proposed \citep[cf.][]{Hochberg87} but we focus here on FWE, the bounding of which is sometimes called \emph{strong error control}, because of its prominence in clinical research \citep{Lehmacher91} and other applications. 

A set of hypotheses $H_1,\ldots,H_k$ is called \emph{closed} if the set  $\{H_1,\ldots,H_k\}$ is closed under intersection.  \citet{Marcus76} introduced a method of testing a closed set of hypotheses $H_1,\ldots, H_k$ that controls the FWE by requiring that there be an $\alpha$-level test of every intersection hypothesis $\cap_{i\in J} H_i$, $J\subseteq \{1,\ldots,k\}$. Let~$\bm{H}$ be the set of all such intersections, and note that closedness of $\{H_1,\ldots,H_k\}$ is equivalent to it being equal to~$\bm{H}$. Beginning with the \emph{global hypothesis} $\cap_{i=1}^k H_i$, Marcus et al.'s procedure tests the non-empty elements of $\bm{H}$ in order of decreasing \emph{dimension} (defined as the maximum number of~$H_i$ being intersected) by their corresponding $\alpha$-level tests,  and proceeding by using the rule that $H\in\bm{H}$ is tested if and only if all elements of $\bm{H}$ contained in $H$ are tested and rejected.  Multistage extension of closed tests is relatively straightforward because of the nature of the hypotheses and the test statistic. It is essentially  a repeated closed testing procedure, performing closed testing of the hypotheses not yet rejected at every stage; see \citet{Tang93} and \citet{Tang99}.

When the set of hypotheses is not closed, Holm's~\citeyearpar{Holm79} step-down procedure  is commonly used for fixed sample size problems. Beginning with a brief review of Holm's~\citeyearpar{Holm79} procedure, Section~\ref{sec:SD} then proceeds to provide a multistage extension, which aims to capture the generality of Holm's procedure for controlling the FWE and to be able to take advantage of the closed testing structure when it exists. A simulation study of the proposed procedure's power and expected sample size is given in Section~\ref{sec:power}, and further applications and relation to the existing litertaure are discussed in Section~\ref{sec:disc}.
 
\section{\large A Multistage Step-Down Procedure}\label{sec:SD}

\citet{Holm79} proposed the following general step-down method of testing $H_1,\ldots,H_k$ that does not assume closedness. Although Holm's procedure has been criticized for lack of power in some settings, it does preserve the FWE without making any assumptions about the structure of the hypotheses or correlations between the individual test statistics, only that for each hypothesis $H_i$ there is a computable $p$-value $\what{p}_i$ such that 
\begin{equation}
P(\widehat{p}_i\le \alpha |H_i)\le \alpha\label{eq:unif}
\end{equation}
 for all $0<\alpha<1$. The $\alpha$-level Holm's procedure proceeds as follows. Compute and order the $p$-values  $\widehat{p}_{i(1)}\le\ldots\le\widehat{p}_{i(k)}$. For $j=1,\ldots, k$, if 
\begin{equation}
\widehat{p}_{i(j)}\ge\alpha/(k-j+1),\label{eq:Holmacc}
\end{equation}
then accept $H_{i(j)},\ldots,H_{i(k)}$;  otherwise, reject $H_{i(j)}$ and move on to stage $j+1$ (if $j<k$). A simple proof that the FWE of Holm's procedure is bounded by $\alpha$ is given in \citet{Lehmann05}. 

As noted above, when closedness exists it is unnecessary to use the step-down correction~(\ref{eq:Holmacc}), or any Bonferroni-type correction for that matter.  However, it is illuminating to now consider how the step-down procedure is related to closed testing procedures. When closedness exists, Marcus et al.'s~\citeyearpar{Marcus76} procedure can be viewed as a special case of Holm's step-down procedure in the following sense. Assume that the $p$-values ``respect'' the closedness in the sense that
\begin{equation}
\label{eq:Holmmono}H_i\subseteq H_j\quad \Rightarrow\quad \what{p}_i\le\what{p}_j.
\end{equation} In this case Holm's procedure will test the elements of $\bm{H}=\{H_1,\ldots,H_k\}$ in order of decreasing dimension (provided we agree to use dimension to break any ``ties''). Assuming that  closedness and~(\ref{eq:Holmmono}) hold, we make two slight modifications of Holm's procedure to utilize these properties. First, upon rejection of~$H_{i(j)}$, we accept any of the remaining hypotheses whose complements are implied by~$H_{i(j)}$, i.e., accept any~$H_{i(j+1)},\ldots, H_{i(k)}$ containing $H_{i(j)}^c$, the complement of $H_{i(j)}$.  This does not change the FWE~$\le\alpha$ bound since~(\ref{eq:Holmmono}) guarantees that the intersection of all true hypotheses, denoted by $G$, is the first true hypothesis tested. Otherwise, if $G$ has already been accepted, then all true hypotheses are subsequently accepted according to this rule.  Next, note that the Bonferroni-type correction in~(\ref{eq:Holmacc}) is now unnecessary, i.e.,  the right-hand-side of~(\ref{eq:Holmacc}) may be replaced by $\alpha$ while maintaining FWE $\le\alpha$, since, letting~$j_G$ denote the rank of the $p$-value associated with~$G$,
\begin{eqnarray*}
\mbox{FWE}&=&P(\mbox{$G$ rejected})\\
&\le&P(\what{p}_{i(j_G)}< \alpha)\\
&\le& \alpha.
\end{eqnarray*} 

We now introduce a multistage generalization of Holm's~\citeyearpar{Holm79} step-down procedure. As with the original version of Holm's procedure, we make no assumptions about the structure of the set of hypotheses $H_1,\ldots, H_k$ to be tested, and the only assumptions are about the family of available tests of the individual hypotheses through their significance levels, following the approach of~\citet{Romano05}. That this procedure satisfies FWE~$\le\alpha$ is proved in Theorem~1.

For each hypothesis $H_i$, assume there is a sequential test statistic $T_{i,n}$, a (non-random) critical value function $C_n(\rho)$ that is non-increasing in $\rho\in(0,1)$, and a set $N$ of possible sample sizes such that 
\begin{equation}
\label{eq:pval} \sup_{\theta\in H_i} P_\theta \left(\sup_{n\in N} [T_{i,n}-C_n(\rho)]\ge 0\right)\le\rho
\end{equation} for all $0<\rho<1$. This requirement is the multistage analog of~(\ref{eq:unif}). The existence of such a family of sequential test statistics may at first seem restrictive, but in many settings there are natural choices for the $T_{i,n}$; see the examples in Sections~\ref{sec:power} and \ref{sec:disc}.  The use of the set $N$ of possible sample sizes allows for the possibilities of fully sequential (e.g.,  $N=\{1,2,3,\ldots\}$) or group sequential (e.g., $N=\{m, 2m, 3m, 4m, 5m\}$ for some $m$) in both the truncated and non-truncated settings.

The $\alpha$-level multistage step-down procedure with no more than $k$ stages is defined as follows. Let $I_1=\{1,\ldots,k\}$, $n_0=0$, and let $|\cdot |$ denote set cardinality. For~$j=1,\ldots, k$:
\begin{enumerate}
\item Sample  up to 
\begin{equation}
\label{eq:seqss}n_j=\inf\left\{n\in N: n> n_{j-1}\quad\mbox{and}\quad\max_{i\in I_j}T_{i,n}\ge C_{n}(\alpha/|I_j|)\right\}.
\end{equation}
\item Order the test statistics $$T_{i(j,1),n_j}\ge T_{i(j,2),n_j}\ge \ldots\ge T_{i(j, |I_j|),n_j}.$$
\item  Reject $H_{i(j,1)},\ldots,H_{i(j,m_j)}$, where 
\begin{equation}
m_j=\max\left\{m\ge 1: \min_{1\le \ell\le m}\left[T_{i(j,\ell),n_j}-C_{n_j}\left(\frac{\alpha}{|I_j|-\ell+1}\right)\right]\ge 0\right\}.\label{eq:numrej}
\end{equation}
\item Stop if $j=k$, if $n_j=\sup N$, or if all remaining hypotheses contain the complement of some rejected hypothesis.  Otherwise, let $I_{j+1}$ be the indices of the remaining hypotheses and continue on to stage~$j+1$.
\end{enumerate} 

\noindent\textbf{Theorem 1.} The multistage step-down procedure satisfies FWE $\le\alpha$. 

\medskip

\noindent\textbf{Proof.}  Let $I$ be the indices of the true hypotheses.  If an error occurs, then for some~$I_j\supseteq I$ and index $\ell$ such that $|I_j|-\ell+1\ge|I|$,
$$\max_{i\in I} T_{i,n_j}\ge C_{n_j}\left(\frac{\alpha}{|I_j|-\ell+1}\right)\ge C_{n_j}(\alpha/|I|),$$ which implies that $$ \max_{i\in I} \sup_{n\in N} \left[T_{i,n}-C_{n}(\alpha/|I|)\right]\ge 0.$$ Then, using the Bonferroni inequality and~(\ref{eq:pval}),
\begin{eqnarray*}
\mbox{FWE}&\le&P\left(\max_{i\in I} \sup_{n\in N} \left[T_{i,n}-C_{n}(\alpha/|I|)\right]\ge 0\right)\\
&\le&\sum_{i\in I}P\left(\sup_{n\in N} \left[T_{i,n}-C_{n}(\alpha/|I|)\right]\ge 0\right)\\
&\le&\sum_{i\in I}\alpha/|I|=\alpha.
\end{eqnarray*}\qed

We point out additionally that the procedure may stopped at Step~4 at any point as long as the remaining hypotheses are accepted, since this action can only serve to decrease the FWE.  This feature may be of use in clinical trial applications; see the last paragraph of Section~\ref{sec:clinical}. 

Theorem~1 holds regardless of the structure of the $H_i$ or the joint distribution of the test statistics~$T_{i,n}$. When the $H_i$ are closed, the above multistage step-down procedure can be modified slightly to take advantage of this additional structure, analogous to the discussion in the second paragraph of this section for the fixed-sample case. To this end, assume that the set $H_1,\ldots,H_k$ is closed, and that the test statistics respect the closedness in the sense that
\begin{equation}
\label{eq:multmono}H_i\subseteq H_j\quad\Rightarrow\quad T_{i,n}\ge T_{j,n}\quad\mbox{for all $n\in N$.}
\end{equation} As in the fixed-sample case, this guarantees that the hypotheses are analyzed by decreasing dimension. First, we modify the multistage step-down procedure by changing Step~3 to:

\bigskip

3.$'$ For $\ell=1,\ldots,m_j$: Reject $H_{i(j,\ell),n_j}$ and accept  any remaining hypotheses containing~$H_{i(j,\ell),n_j}^c$.

\bigskip

\noindent Next, we replace the fractions of $\alpha$ in~(\ref{eq:seqss})-(\ref{eq:numrej}) by $\alpha$. These modifications do not cause violation of FWE~$\le\alpha$ by the same proof given in the second paragraph of this section, proving the following.

\bigskip

\noindent\textbf{Theorem 2.} If the set of hypotheses $H_1,\ldots,H_k$ is closed and the test statistics $T_{i,n}$ satisfy~(\ref{eq:multmono}), then the multistage step-down procedure with Step~3 replaced by~3$'$ and $\alpha$ as the argument of $C_n$ in~(\ref{eq:seqss})-(\ref{eq:numrej}) satisfies FWE~$\le\alpha$.

\section{\large Power and Expected Sample Size}\label{sec:power}

The Holm step-down procedure's attractive quality is its generality, i.e., no assumptions about the structure of the hypotheses $H_1,\ldots,H_k$ or the individual test statistics, other than~(\ref{eq:unif}), are necessary. This generality is provided by the Bonferroni-type adjustment~(\ref{eq:Holmacc}) which also can cause Holm's procedure to be conservative, in terms of FWE and power, relative to procedures that take into account correlations between the individual test statistics. This conservativeness is shared by the multistage step-down procedure because of its use of an analogous step-down rule~(\ref{eq:numrej}). However, as pointed out above, the utility of either the multistage or fixed sample step-down procedure lies in cases where such correlations are difficult to model.

\begin{table}[htdp]
\caption{\;A 3-Endpoint Trial}
\begin{center}
\begin{tabular}{l|cccccc}
Procedure&$(\mu_1,\mu_2,p)$&$EM$&$P(\mbox{rej.~$H_1$})$&$P(\mbox{rej.~$H_2$})$&$P(\mbox{rej.~$H_3$})$&FWE\\
\hline
H&&105&1.7\%&1.7\%&0.8\%&4.0\%\\
Mult&$(0,0,.5)$&104.7&1.6\%&1.6\%&1.8\%&4.9\%\\
MultH&&104.6&1.5\%&1.5\%&2.0\%&4.8\%\\\hline
H&&105&2.3\%&2.3\%&76.0\%&4.4\%\\
Mult&$(0,0,.75)$&98.4&1.7\%&1.7\%&79.9\%&3.2\%\\
MultH&&98.3&2.1\%&2.1\%&80.7\%&4.2\%\\\hline
H&&105&2.5\%&95.7\%&2.1\%&4.4\%\\
Mult&$(0,.65,.5)$&96.9&1.6\%&94.5\%&1.9\%&3.4\%\\
MultH&&96.9&2.2\%&94.6\%&2.9\%&4.9\%\\\hline
H&&105&4.3\%&82.9\%&83.9\%&4.3\%\\
Mult&$(0,.5,.75)$&92.8&1.5\%&76.3\%&80.3\%&1.5\%\\
MultH&&92.3&3.2\%&79.9\%&85.2\%&3.2\%\\\hline
H&&105&83.4\%&83.4\%&3.8\%&3.8\%\\
Mult&$(.5,.5,.5)$&93.5&76.3\%&76.3\%&1.8\%&1.8\%\\
MultH&&93.1&79.6\%&79.6\%&2.7\%&2.7\%\\\hline
H&&105&70.9\%&70.9\%&86.8\%& NA\\
Mult&$(.4,.4,.75)$&89.9&55.3\%&55.3\%&80.2\%& NA\\
MultH&&89.3&64.7\%&64.7\%&85.4\%& NA\\\hline
H&&105&88.6\%&88.6\%&90.0\%&NA\\
Mult&$(.5,.5,.75)$&87.2&76.4\%&76.4\%&80.0\%& NA\\
MultH&&86.1&84.4\%&84.4\%&87.0\%& NA\\\hline
\end{tabular}
\end{center}
\label{table:3end}
\end{table}%

Consider a multistage step-down procedure with maximum sample size $n=\max N$. Relative to the fixed-sample Holm step-down procedure of size $n$, the multistage procedure will have a reduction in expected sample size, provided the set $N$ is chosen reasonably. But, by the Neyman-Pearson lemma, the power of the multistage procedure for rejecting a given false hypothesis cannot exceed the power of the fixed-sample Holm procedure. However, as shown by the following simulation study, this loss in power is usually slight while the savings in expected sample size tends to be substantial.

Consider a three-endpoint clinical trial where two of the endpoints concern continuous data and the third concerns probability of a certain binary outcome. For example, let the data be $\bm{X}_{i}=(X_{i1}, X_{i2}, X_{i3})$, $i=1,2,\ldots$, where for $j=1,2$, the $X_{ji}$  are i.i.d.~normal random variables with unknown mean $\mu_j$ and variance~1, and the $X_{3i}$ are independent Bernoulli random variables where $p=P(X_{3i}=1)$ is unknown. Suppose the $\bm{X}_i$  represent clinical treatment outcomes for three endpoints of interest, and it is desired to test efficacy of the treatment in the form of the three one-sided null hypotheses $$H_1:\mu_1\le 0,\quad H_2:\mu_2\le 0,\quad H_3:p\ge 1/2.$$ In cases such as this, the correlation between the components of $\bm{X}_{i}$ is likely to be unknown or difficult to model. In the following simulation study we compare the performance of the multistage step-down procedure with two other procedures in the three cases of independent, positively correlated, and negatively correlated components of $\bm{X}_i$. Table~1 contains the results for the independent case; the other two are discussed below.  Whatever the correlation between the components, the three procedures evaluated are equally applicable since they do not depend on the correlation structure of the individual hypotheses. For Holm's step-down procedure, we use standard $\alpha=.05$-level likelihood ratio tests of $H_1, H_2, H_3$ of size $n=35$ to have power around 90\% at $(\mu_1,\mu_2,p)=(.5,.5,.75)$. For the multistage step-down procedure, we use the same test statistics in one-sided group sequential tests with $N=\{26,29,35\}$ and use a normal approximation for $\sum_i X_{3i}$ to compute $C_n(\rho)$ to satisfy~(\ref{eq:pval}). Table~1 contains the expected sample size, probability of rejecting each $H_i$, and FWE (when $\cup_{i=1}^3 H_i$ is true) for the Holm procedure (denoted by H) and the multistage step-down procedure (denoted by MultH). To see the effects of the step-down rule~(\ref{eq:seqss})-(\ref{eq:numrej}), we also include the multistage test (denoted by Mult) identical to MultH but with $\alpha$ divided by~$k=3$ in place of the larger fractions of $\alpha$ in~(\ref{eq:seqss})-(\ref{eq:numrej}); see the first paragraph of the Section~\ref{sec:k}.   Each entry in Table~1 is computed from 50,000 simulation runs. The last six FWE entries are marked NA (not applicable) because none of the null hypotheses are true for those parameter values. The multistage procedures Mult and MultH show substantial savings in expected size sample over the Holm procedure's fixed sample size of 105 at significant deviations from the ``worst-case'' null $(\mu_1,\mu_2,p)=(0,0,.5)$. The expected sample size of Mult and MultH are nearly identical, the former being somewhat larger due to its larger critical values in~(\ref{eq:seqss})-(\ref{eq:numrej}). For the same reason MultH has slightly higher power than Mult; in particular, see the last six rows of Table~1. Although the power of MultH is lower than H due to its multiple looks, this difference is slight, usually within a few percentage points. This relative relationship does not change when the components of $\bm{X}_i$ are correlated, it simply tends to decrease slightly in magnitude when positively correlated, and increase when negatively correlated. For example, when $(\mu_1,\mu_2,p)=(0,.5,.75)$ and the two normal components have a correlation coefficient of $.75$, $P(\mbox{reject $H_1$})$ increases to 4.5\%, 2.0\%, and 3.6\% for H, Mult, and MultH, respectively, while the power $P(\mbox{reject $H_2$})$ decreases to 80.7\%, 73.4\%, and 77.0\%, respectively.  Here Mult and MultH have expected sample size of 94.7 and 94.2.

\section{\large Applications and Discussion}\label{sec:disc}

\subsection{Sequential $k$-Hypothesis Testing}\label{sec:k}
A straightforward application of~(\ref{eq:pval}) to sequential testing of $k$ null hypotheses is to use Bonferroni's inequality so that the $k$ hypotheses can be treated separately by setting~$\rho=\alpha/k$ in~(\ref{eq:pval}). This approach to sequential multiple testing has been taken by many authors. \citet{Paulson64} noticed that further sample size savings might be possible by eliminating (rejecting) some hypotheses during the course of the experiment, similar to the test Mult in the preceding section. We have refined the multistage rejection of hypotheses by using Holm's step-down procedure to sharpen the Bonferroni bounds. In fact, when the set of hypotheses is closed, a slight modification of our multistage test can dispense with Bonferroni bounds, as shown in Theorem~2.

Sequential multiple hypothesis testing dated back to~\citet{Sobel49} in deciding which of three simple hypotheses $H_1: \theta=-d$, $H_2:\theta=0$, or $H_3:\theta=d$ is true about a normal mean $\theta$, for a fixed value $d>0$. This is basically a classification problem. The Sobel-Wald test combines  two sequential probability ratio tests~(SPRTs) for different pairs of the three hypotheses, the comparison of $H_1$ versus $H_3$ being superfluous. \citet{Armitage50} generalized the Sobel-Wald problem to $k$~hypotheses, corresponding to an error matrix $\alpha_{ij}=P_i(\what{i}=j)$ for $i\ne j$, where $P_i$ denotes the probability measure under $H_i$ and $\what{i}$ is the hypothesis chosen by the test. Armitage's test combines the corresponding ${k\choose 2}$ SPRTs for the $k$ hypotheses, which leads to stopping boundaries with slope $\pm 1/2$ in the $(n,S_n/d)$-plane, where $S_n=\sum_{i=1}^n X_i$ is the sum of the first $n$ i.i.d. normal observations. \citet{Simons67} considered a generalization of a special case of Armitage's test for $k=3$ in which the stopping boundaries' slopes can be chosen arbitrarily. Whereas the Sobel-Wald, Armitage, and Simons tests stop and decide on a hypothesis $H_j$ when all the component SPRTs simultaneously prefer $H_j$ to all other alternatives, \citet{Lorden72,Lorden76} introduced various multistage tests that decide on $H_j$ only when all other hypotheses can be rejected, based on generalized likelihood ratios. Lorden's work was extended to composite hypotheses by \citet{Pavlov88}. \citet{Eisenberg91} gives a detailed summary of these problems. 
 
\citet{Paulson63} introduced another generalization of the Sobel-Wald test, considering $k\ge 2$ intervals $(-\infty,\theta_1)$, $(\theta_1,\theta_2),\ldots$, $(\theta_{k-1},\infty)$ and testing sequentially to which interval $\theta$ belongs. In its symmetric form, Paulson's test stops the first time the interval $(u_n,v_n)$ is contained in one of the intervals $(-\infty,\theta_1+\delta)$, $(\theta_{k-1}-\delta,\infty)$, or $(\theta_i-\delta,\theta_{i+1}+\delta)$, $i=1,\ldots,k-2$, where  $\delta>0$ is a chosen parameter and
\begin{eqnarray}
u_n=\max_{1\le m\le n}\wt{u}_m=\max_{1\le m\le n} (S_m/m-\delta/2-A/m)\label{eq:un}\\
v_n=\min_{1\le m\le n}\wt{v}_m=\min_{1\le m\le n} (S_m/m+\delta/2+A/m)
\end{eqnarray} in which $A>0$ is a critical value that can be chosen to give desired coverage probability.  Paulson's procedure can be viewed as a special case of the multistage step-down procedure, as follows. Defining $H_0^{(i)}: \theta=\theta_i-\delta$ and $H_1^{(i)}: \theta=\theta_i$ ($1\le i\le k-1$), the one-sided SPRT of $H_0^{(i)}$ versus~$H_1^{(i)}$ stops sampling and rejects $H_0^{(i)}$ if
\begin{equation}
S_n-n[(\theta_i-\delta)+\theta_i]/2\ge A,\label{eq:SPRTrr}
\end{equation} for some critical value $A>0$. Dividing both sides of~(\ref{eq:SPRTrr}) by $n$ and rearranging terms gives 
$$A/n\le S_n/n-(\theta_i-\delta/2)=S_n/n-\delta/2-(\theta_i-\delta),$$ which by~(\ref{eq:un}) is equivalent to $\wt{u}_n\ge\theta_i-\delta$. Similarly, the one-sided SPRT of $\wt{H}_0^{(i)}: \theta=\theta+\delta$ versus $H_1^{(i)}$ stops sampling and rejects $\wt{H}_0^{(i)}$ if $\wt{v}_n\le\theta_i+\delta$. Hence running the multistage step-down procedure until a coherent classification is made is precisely Paulson's procedure. In the small-probability event that no coherent classification is made -- say, if at some stage a hypothesis is rejected containing the complement of a previously rejected hypothesis -- classification based on $S_n/n$ or randomization can be used, as \citet{Paulson63} suggests.

Although we have used $H_0^{(i)}$ and $\wt{H}_0^{(i)}$ to denote null hypotheses for combining SPRTs above, there is no natural notion of a ``null'' hypothesis in the actual classification problem.  We could have populated our list of hypotheses to test in any way that suited our needs. Thus it is also perhaps interesting to consider what closed testing has to say about the Sobel-Wald-Paulson problem. In particular, for the case $k=3$ in the Paulson problem, let $H_1:\theta<\theta_1$, $H_2:\theta>\theta_1$, $H_3:\theta<\theta_2$, and $H_4:\theta>\theta_2$.  The set $\{H_1,H_2,H_3,H_4\}$ is not closed, but adding $H_5=H_2\cap H_3=(\theta_1,\theta_2)$ to the list of hypotheses completes its closure $\bm{H}=\{H_1,H_2,H_3,H_4,H_5\}$. Since all intersections of dimension~3 or higher are empty, the closed testing principle suggests beginning by testing the hypotheses of dimension~2, namely $H_1=H_1\cap H_3=(-\infty,\theta_1)$, $H_4=H_2\cap H_4=(\theta_2,\infty)$, and $H_5=(\theta_1,\theta_2)$, which is of course the original Paulson problem with $k=3$. Hence it seems that the closed testing principle does not give any new insight into the Paulson problem. 

A closely related problem to sequential $k$-hypothesis testing is selecting the one of $k$ normal populations with the largest mean. \citet{Bechhofer54} considered this problem when the variance of the observations is known, and proposed a fixed sample procedure that compares the sample means of the individual populations.  For unknown variance, \citet{Bechhofer54b} proposed a two-stage procedure, and \citet{Robbins68} proposed a sequential procedure with improved efficiency, also based on sample means. The special case of when the mean is an integer was considered by \citet{Robbins70}, and later generalized by \citet{McCabe73}.  \citet{Robbins70} introduced the notion of ``distinguishability'' of a family of populations, and \citet{Khan73} studied the asymptotic efficiency of stopping rules that distinguish within such families. \citet{Mukhopadhyay83} proposed likelihood-based methods for the largest mean problem, and used Khan's results to show asymptotic efficiency.  Likelihood-based methods were shown to be useful in a number of related selection problems as well; see \citet{Mukhopadhyay94}. Further references are given in Section~2 of \citet{Chan05}.

\subsection{Multiple Endpoint Clinical Trials}\label{sec:clinical}
The multistage step-down procedure provides a general method of testing multiple endpoints in clinical trials.  The adaptive rejection times~(\ref{eq:seqss}) and rejection rule~(\ref{eq:numrej}) have the effect of adaptively ``dropping'' (i.e., rejecting) hypotheses when enough information has accumulated to do so, to focus on the statistically most interesting endpoints. As discussed above, when closedness exists, closed testing methods should be used since they are are in general more powerful than step-down methods (simulations verifying this were conducted by \citet{Lehmacher91} and \citet{Tang97}) because they forgo the need for Bonferroni-type corrections, such as in~(\ref{eq:Holmacc}), (\ref{eq:seqss}), and (\ref{eq:numrej}). The multistage step-down method could be useful in cases where closedness does not exist. For example, in clinical trials for AIDS treatments, it is common \citep[e.g.,][]{Fischl87} to have multiple endpoints of both the continuous and categorical types, like CD4 (T-cell) level, which is commonly modeled as a normal random variable, and the binary indicator of opportunistic infectious disease like a cold, modeled as a Bernoulli random variable. Moreover, if one or some subset of the endpoints is of primary interest, the multistage step-down procedure can be used as a pilot or screening phase with the option of immediately stopping and proceeding to secondary testing when one of the primary hypotheses is rejected. That this does not increase the FWE is pointed out following the proof of Theorem~1. The multistage step-down procedure provides a general framework that can be applied to multiple testing in these clinical trials. 

\section*{\large Acknowledgments}

Bartroff's work was supported by the Borchard Foundation and grant DMS-0907241 from the National
Science Foundation. Lai's work was supported by grant DMS-0805879 from the National
Science Foundation.


\end{document}